\documentclass[aps,pre,twocolumn,superscriptaddress,preprintnumbers,amsmath,amssymb]{revtex4}
\usepackage{amsmath}
\usepackage{graphicx}
\usepackage{dcolumn}
\usepackage{bm}
\usepackage{float}
\usepackage{color}

\begin{document}

\title{Time-resolved characterization of the formation of a collisionless shock}


\author{H. Ahmed}
\affiliation{Centre for Plasma Physics, School of Mathematics and Physics, Queen's University of Belfast, Belfast BT7 1NN, United Kingdom}

\author{M. E. Dieckmann}
\affiliation{Centre for Plasma Physics, School of Mathematics and Physics,  Queen's University of Belfast, Belfast BT7 1NN, United Kingdom}

\author{L. Romagnani}
\affiliation{LULI, \'Ecole Polytechnique, CNRS, CEA, UPMC, Palaiseau, France}

\author{D. Doria}
\affiliation{Centre for Plasma Physics, School of Mathematics and Physics,  Queen's University of Belfast, Belfast BT7 1NN, United Kingdom}

\author{G. Sarri}
\affiliation{Centre for Plasma Physics, School of Mathematics and Physics,  Queen's University of Belfast, Belfast BT7 1NN, United Kingdom}

\author{M. Cerchez}
\affiliation{Institute for Laser and Plasma Physics, University of D\"usseldorf, Germany}

\author{E. Ianni}
\affiliation{Centre for Plasma Physics, School of Mathematics and Physics,  Queen's University of Belfast, Belfast BT7 1NN, United Kingdom}

\author{I. Kourakis}
\affiliation{Centre for Plasma Physics, School of Mathematics and Physics,  Queen's University of Belfast, Belfast BT7 1NN, United Kingdom}

\author{A. L. Giesecke}
\affiliation{Institute for Laser and Plasma Physics, University of D\"usseldorf, Germany}

\author{M. Notley}
\affiliation{Central Laser Facility, Rutherford Appleton Laboratory, Chilton, Oxfordshire OX11 0QX, United Kingdom}

\author{R. Prasad}
\affiliation{Centre for Plasma Physics, School of Mathematics and Physics,  Queen's University of Belfast, Belfast BT7 1NN, United Kingdom}

\author{K. Quinn}
\affiliation{Centre for Plasma Physics, School of Mathematics and Physics,  Queen's University of Belfast, Belfast BT7 1NN, United Kingdom}

\author{O. Willi}
\affiliation{Institute for Laser and Plasma Physics, University of D\"usseldorf, Germany}

\author{M. Borghesi}
\affiliation{Centre for Plasma Physics, School of Mathematics and Physics,  Queen's University of Belfast, Belfast BT7 1NN, United Kingdom}
\affiliation{Institute of Physics of the ASCR, ELI-Beamlines Project, Na Slovance 2, 18221 Prague, Czech Republic}


\date{\today}

\begin{abstract}
We report on the temporally and spatially resolved detection of the precursory stages that lead to the formation of an unmagnetized, supercritical collision-less shock in a laser-driven laboratory experiment. The measured evolution of the electrostatic potential associated with the shock unveils the transition from a current free double layer into a symmetric shock structure, stabilized by ion reflection at the shock front. Supported by a matching Particle-In-Cell simulation and theoretical considerations, we suggest that this process is analogeous to ion reflection at supercritical collisionless shocks in supernova remnants.
\end{abstract}

\maketitle
Collision-less shocks represent particularly intriguing phenomena in plasma physics, due to their implications in a broad range of physical scenarios, extending from laboratory-based laser-plasma experiments to astrophysics. In the latter case, particular attention has been devoted to shock waves generated during the propagation of a supernova remnant (SNR) blast shell into the interstellar medium (ISM), since they are thought to be the dominant source of galactic high energy cosmic rays \cite{Koyama,Raymond,Abdo,Helder,Lieberman}. In this case, non-collisionality is guaranteed by the low particle collision frequency in the ISM \cite{Ferriere}; the dynamics of SNR shocks is expected to be dominated by electromagnetic fields, thus setting stringent limits on their speed and stability. However, despite the considerable number of recent observations of such structures, the intrinsic difficulty in directly probing the plasma conditions around the SNRs has left the debate upon their generation mechanism and dynamics still open in the scientific community. A possible solution to this impasse might be provided by studying small-scale reproductions of these phenomena in laser-based laboratory experiments. Within this framework, promising results have indeed already been obtained \cite{Romagnani,Gregori,Kuramitsu}, especially concerning the stationary stage of these structures. Nonetheless, a full understanding of the dynamics of a collision-less shock would also require a detailed characterization of the transient phase in which it is formed, a regime that has hitherto eluded experimental detection.

Employing a time-resolved proton imaging technique \cite{Sarri2}, we present here the first experimental observation of the precursory stages that lead to the generation of a supercritical electrostatic collision-less shock at the boundary of a blast shell of laser-ablated plasma expanding into a dilute ambient medium. By following the temporal evolution of the propagation speed and of the profile of the associated electrostatic potential, it has been possible to distinguish the intermediate steps that let an initially freely expanding structure, similar to a current free double layer (CFDL) \cite{DoubleLayer,Singh}, evolve into a forward shock propagating at a supercritical speed (observed Mach number of the order of 4). A matching Particle-In-Cell (PIC) simulation allowed us to reproduce the transformation of the plasma contact boundary by collision-less electrostatic processes
into a forward and reverse shock on spatio-temporal scales that are similar to those in the experiment. The late-time behavior of such a shock,
which is revealed by the simulation, shows that the shock-reflected ions locally heat the plasma in the close vicinity of the shock (foreshock) and 
thus increase the effective ion-acoustic speed, if compared to the one of the unperturbed upstream plasma. This heating is not related to beam-driven instabilities but to the electrostatic potential associated to the shock-reflected ion beam and the increased ion-acoustic speed directly ahead of
the shock contributes to its stabilization. This effect  may also help stabilizing supernova remnant (forward) shocks like that of RCW86 \cite{Helder} by
heating up the ISM plasma directly ahead of the shock.

\begin{figure}[!t]
\includegraphics[width=\columnwidth]{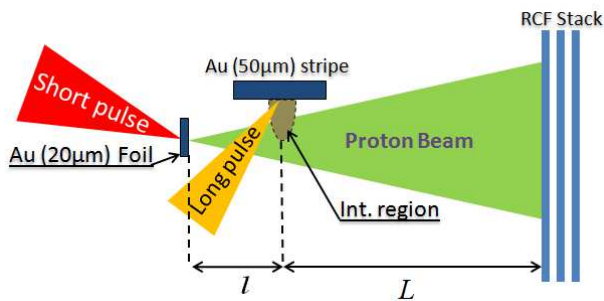}
\caption{Sketch of the typical time of flight experimental arrangement of the proton imaging technique. Here $L=38$ mm and $l=4$ mm, giving an intrinsic geometrical magnification of $M \approx(l+L)/l \approx 10.5$. The long pulse is incident at $45^\circ$ to the target's normal.\label{fig1}}
\end{figure}

In the experiment, carried out at the VULCAN laser facility at the Rutherford Appleton Laboratory \cite{VULCAN}, a 1 ns, 120 J laser pulse (Long pulse in Fig. \ref{fig1}) was focussed, down to a focal spot of 50 $\mu$m radius, onto a 50 $\mu$m thick gold foil, resulting in a peak intensity of the order of $I_L \approx 10^{15}$ Wcm$^{-2}$. The gold foil was surrounded by nitrogen gas at a pressure of $10^{-1}$ mbar. A thin layer of hydrocarbon and vapor contaminants (thickness of the order of tens of nm), as resulting from deposition of moisture and impurities of the low-density gas \cite{Gitomer,Tan}, will be present on the gold foil surfaces.
The laser-foil interaction generates a warm and dense plasma radially expanding in the form of a rarefaction wave \cite{Sarri1}. This plasma will be constituted of thermally distributed heavy ions with a temperature of the order of the keV (gold) plus a component of faster lighter ions from surface contaminants, with an average energy per nucleon of tens of keV \cite{Gitomer,Tan}. Hydrodynamic simulations (carried out with the HYADES code \cite{Hyades}) indicate that this dense plasma expands into a more rarefied background plasma induced by the full ionization of the nitrogen gas by secondary X-ray emission from the solid target \cite{Dean}, with an average electron temperature and density of 650 eV and $10^{16}$ cm$^{-3}$, respectively.
Such an experimental configuration resembles the one already adopted in Ref. \cite{Romagnani}, although with a much higher ambient plasma density. The plasma evolution was monitored by temporally and spatially resolved proton radiographs with a resolution of approximately a picosecond and a few microns, respectively \cite{Sarri2}. The proton beam was externally generated during the interaction of a secondary laser beam (intensity $I_{L2} \approx 10^{19}$ Wcm$^{-2}$, Short pulse in Fig. \ref{fig1}) with a 20 $\mu$m thick gold foil. After having traversed the plasma, the proton beam was recorded on a stack of RadioChromic Films (RCF) \cite{Dempsey}. The adjustable delay between the two laser beams, allowed us to temporally scan the plasma evolution, until the temporal window in which a collisionless shock started to form was found.

\begin{figure}[!t]
\includegraphics[width=\columnwidth]{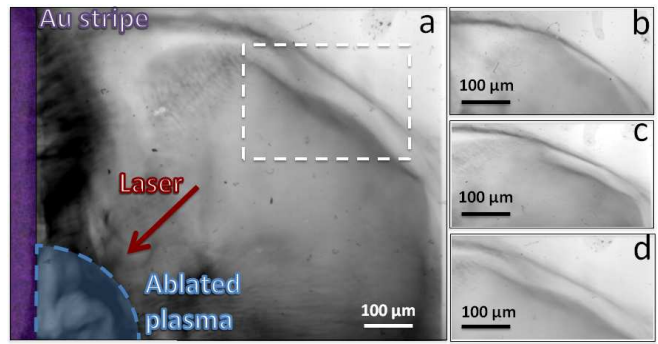}
\caption{Two-dimensional proton radiographs: (a) Proton radiograph of the interaction of a nanosecond laser pulse (red arrow) with a 50 $\mu$m gold foil (left purple rectangle) corresponding to 170ps after the arrival of the long pulse (see Fig. \ref{fig1}). Zooms of the proton radiographs of the region highlighted by the dashed white rectangle in the frame (a) for different proton energies of approximately 11.5 (frame (b)), 10 (frame (c)), and 9 (frame (d)) MeV. These energies correspond to a probing time of 150, 160 and 170 picoseconds after the start of the interaction, respectively.\label{fig2}}
\end{figure}

\begin{figure}
\includegraphics[width=\columnwidth]{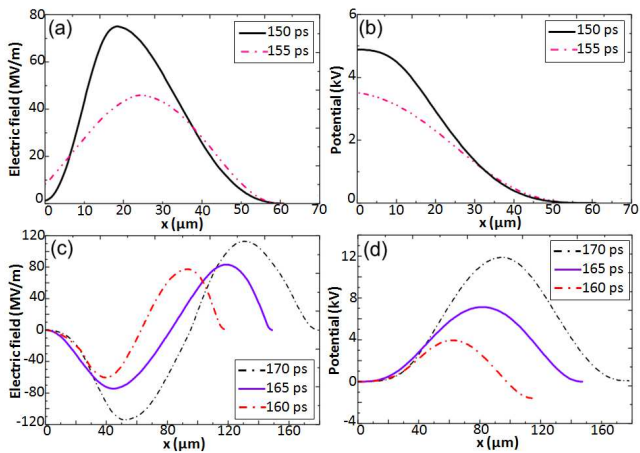}
\caption{The electric field and potential: Experimental electric field distribution (a,c) and associated potential (b,d) during the transition between the CFDL and the proto-shock. The experimental error on the electric field and position are of the order of 8\% and 5 $\mu$m respectively. The profiles refer to sequential stages of the same event, and are shown in separate graphs for ease of clarity.\label{fig3}}
\end{figure}

As an example, Fig. \ref{fig2}.a depicts a typical proton radiograph of the interaction, in which the probing proton beam were timed so to traverse the plasma region under interest 170 ps after the arrival of the long pulse on the gold target. The image shows two main regions of proton deflections. A turbulent pattern is visible in proximity to the gold target surface (labeled by `Ablated plasma' in Fig. \ref{fig2}.a) and it is induced by the electrostatic and magnetic fields associated with the bulk region of the expanding gold plasma (similarly to what discussed in Ref. \cite{Romagnani}). Ahead of this, at $\sim$ 1 mm from the target, an approximately circular pattern with a radius of curvature of $\sim$ 0.9 mm is also visible. As a rule of thumb, it is worth recalling that darker grey colors indicate a higher proton deposition (proton accumulation) whereas lighter grey colors indicate a region of proton depletion. Bearing this in mind, this structure is made of two dark lines of proton accumulation separated by a lighter stripe of proton depletion. Due to the multi-frame capability of the proton imaging technique \cite{Sarri2}, it has been possible to follow the temporal evolution of this structure, in a single laser shot, within a temporal window of a few tens of ps and with a temporal resolution of the order of a ps.
As an example, Figs. \ref{fig2}.b, \ref{fig2}.c and \ref{fig2}.d depict proton radiographs of the same region (highlighted by a dashed white rectangle in Fig. \ref{fig2}.a) at 150, 160 and 170 ps after the arrival of the long pulse, respectively. The structure is seen to evolve, while roughly preserving its overall shape. Meanwhile, the propagation velocity is seen to decrease from an initial value of the order of $v_1 \approx 2\times10^6$ m/s down to a velocity of approximately $v_2 \approx8\times10^5$ m/s within a few tens of ps. The ambient plasma parameters inferred from HYADES simulations indicate an ion-acoustic velocity of $c_s\approx 2\times10^5$ m/s implying that these velocities would correspond to a Mach number of $M_1\approx 10$ and $M_2 \approx 4$, respectively. We infer that this shock-like structure is collisionless and predominantly electrostatic. A collisional shock would have a thickness comparable to the ion-ion mean free path (i.e. 3 cm for our experimental parameters). A magnetized shock would have a thickness of the order of the ion-gyroradius. By assuming the nitrogen to be fully ionized (as indicated by hydrodynamic simulations), the ion-gyroradius can be expressed as: $r_B \approx m_i v_2 / (q_i B)$ being $q_i = 7e$ and $m_i = 14 m_p$ the nitrogen ion charge and mass, respectively. A ion-gyroradius of the order of 100 microns (measured width of the observed structure) would imply an instability-driven magnetic field of the order of $B\approx 2$ MG, which is unrealistically large in this plasma. The electrostatic nature of the shock, which is also in agreement with previous experimental \cite{Romagnani} and numerical \cite{Sarri1,Sarri3} work, allows us to quantitatively extract the temporal evolution of the electric field distribution associated with the shock using Eq. (5) in Ref. \cite{Sarri2} (see Figs. \ref{fig3}(a,c)).
The electric field evolves from a bell-like shape (Fig. \ref{fig3}.a) to a symmetric bipolar structure, passing through an asymmetrically distributed intermediate state (Fig. \ref{fig3}.c), within 20ps, while keeping an amplitude of the order of 80 - 100 MV/m.

This transition is a clear signature of the precursory stages of the formation of a shock, according to the following dynamics. The propagation of a dense plasma into a more dilute one induces a density gradient at the blast shell front. Since the thermal speed of electrons is larger than that of equally hot ions by a factor ${(m_i / m_e)}^{1/2} \ge 40$, the electrons diffuse out into the ambient plasma and an ambipolar electrostatic potential is set. This structure is called a CFDL \cite{DoubleLayer,Singh} if the electric field is stationary in its rest frame.
The progressive pile-up of ambient ions by this potential introduces a local density maximum at the blast shell front. Initially the density distribution and, thus, the potential on both sides of this maximum are asymmetric; the density jump is larger towards the ambient plasma. The continuing ion pile-up forces the electric field pulse to evolve into a symmetric profile. As soon as the potential is strong enough to reflect the incoming ambient and blast shell ions a forward and a reverse shock form. We focus on the potential jump towards the ambient medium. A potential jump of approximately 4 kV at $t$=150 ps [Fig. \ref{fig3}(b)] can not decelerate the ambient nitrogen ions, which move at 10 $c_s$ relative to the blast shell front, even if they are fully ionized. The blast shell front is expanding freely. The potential [Fig. \ref{fig3}(d)] at $t$=170 ps has increased to 12 kV and the structure's speed relative to the ambient plasma has decreased to $\approx 4 c_s$. This potential is now sufficient to reflect ambient nitrogen ions in the reference frame of the expanding structure provided their charge state is $\ge 4+$. The measured Mach number $M_2 \approx 4$ implies that a supercritical forward shock is forming.

In order to better elucidate the dynamics discussed above, a PIC simulation has been performed with initial conditions based on those of the ambient medium in the experiment. The convergence of the experimentally measured electric field distribution to a symmetric bipolar pulse motivates the study of the collision of two symmetric plasma clouds, each with an electron density of $n_0 = 3.5 \times 10^{16}$cm$^{-3}$ counterpropagating with a relative speed $v_c \approx 5.5 c_s$. The electron and ion temperature is set to $T = 650$ eV. Electrons and ions in each cloud are resolved by 4464 computational particles. The left plasma cloud occupies the interval $-L/2 < x < 0$ and the right plasma cloud occupies the interval $0 < x < L/2$. The box length $L = 3.6$ mm is resolved by 3000 cells and we adopted periodic boundary conditions. The plasma
evolution is followed for $T_\mathrm{sim} =  140$ ps.
\begin{figure}\suppressfloats
\includegraphics[width=\columnwidth]{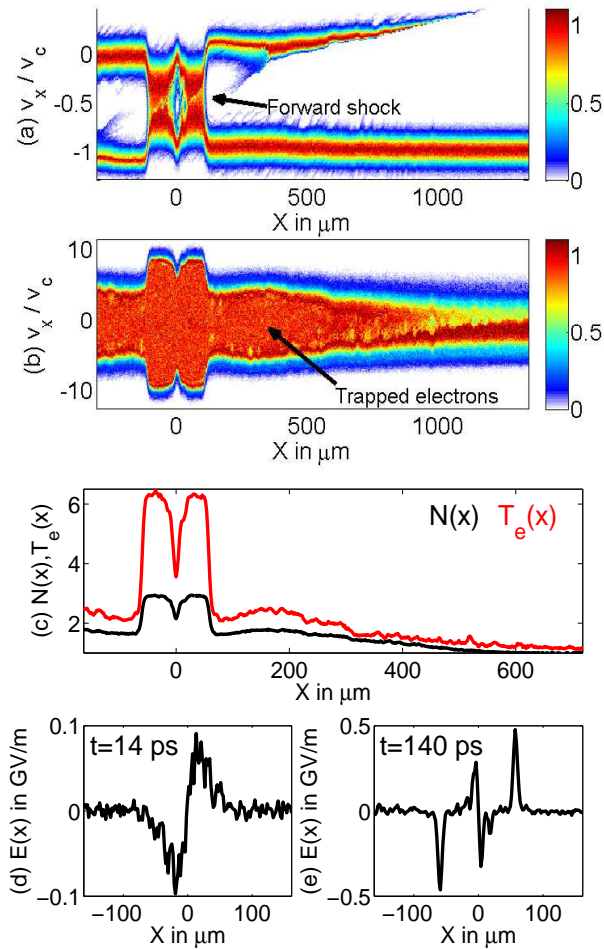}%
\caption{The simulation results: Simulation plasma at $t=140$ ps. (a) shows the ion phase space density on a linear scale. Shocks are located at $x\pm 60 \mathrm{\mu m}$ and an ion phase space hole at $x\approx 0$. The electron phase space density is shown on a linear scale in (b). (c) shows the distributions of the ion density $N(x)$ and electron thermal energy density $T_e(x)$ normalized to those of the upstream, unperturbed plasma. (d) and (e) show the electric field profiles close to $x\approx 0$ at two times.\label{fig4}}
\end{figure}

The interpenetration of the two plasma clouds results in a density pile-up which sets a net positive electrostatic field at the interface. In such an overlap layer, the ions are thus slowed down, increasing their local density beyond 2$n_0$. Figure \ref{fig4}.d reveals at this time a bipolar electric field. Its extent ($\approx 100 \mathrm{\mu m}$) and peak amplitude of $\approx 80$ MV/m are practically identical to those observed in the experiment (compare Fig. \ref{fig4}.d with Fig. \ref{fig3}.c). This structure develops during 14ps, comparable to the time in which the unipolar electric pulse is transformed into a bipolar one in the experiment. Later in time ($t=140$ ps), several electric field peaks appear (see Fig. \ref{fig4}.e). An ion phase space hole \cite{Bujarbarua}  is responsible for the stationary bipolar electric field at $x\approx 0$. Identical shocks are located at $x \pm 60 \mathrm{\mu m}$ and confine the downstream region. We can distrínguish three plasma regions in Fig. \ref{fig4}.c and $x>0$: downstream ($x < 60 \mu$m), foreshock ($60 < x < 600 \mu$m) and upstream ($x > 600 \mu$m). Two unipolar electric pulses, each with a peak electric field of $\sim$ 0.5 GV/m and propagating at a speed of $v_s = \pm 3.4 c_s$ in the upstream reference frame, are generated. A potential jump of +5.2 kV at the
forward shock as we go from the foreshock to the downstream region and a gradual decrease of the peak foreshock potential from 
+1.3 kV at $x \approx  180 \mu m$ to the reference value 0 kV at the far upstream region are thus set, implying a cumulative potential between the downstream and upstream region of 6.5 kV. We have computed these values from an integration of the
electrostatic field (not shown). This potential can reflect the computational ions that move with the relative speed 3.4$c_s$ and have a mass equal to 250 electron masses, generating a beam with density 0.7$n_0$ (See Fig. \ref{fig4}.a). The foreshock potential relative to the 
upstream accelerates incoming electrons towards the shock. The electron core population is at rest at 
$x\approx 1200 \mathrm{\mu m}$ and Fig. \ref{fig4}.b shows that these electrons have been accelerated to about -5$v_c$
at $x\approx 200\mu m$. They gain a kinetic energy of about 1-2keV. This electron acceleration is entirely due to the ambipolar electric
fields arising from the ion density gradient. The potential jump at $x\approx 60 \mathrm{\mu m}$ enforces a reduction of the typical electron speeds as we go from the downstream to the foreshock region, but we find fast electrons with $v\approx 5v_c$ that leak into the foreshock (Fig. \ref{fig4}.b). The electron mixing in the foreshock raises the thermal energy density (Fig. \ref{fig4}.c) to about 1/3 of the downstream one and the ion-acoustic speed is raised to $c_s^* = \sqrt{2} c_s$, where $c_s$ is the ion-acoustic speed of the unperturbed upstream plasma. The effective Mach number of the shock is decreased to $M_s^* = 3.4 / \sqrt{2} \approx 2.4$. This reduction of the effective Mach number might seem modest; however, the electron's thermal spread in the foreshock depend on the relative speed between the electrons leaking from the downstream region and the incoming upstream electrons, which have been accelerated towards the shock by the potential between the foreshock and the far upstream. It is thus independent of the upstream electron temperature. This effect may thus be much stronger at SNR shocks, where the downstream electron temperature exceeds by far that of the far upstream (ISM) plasma. 

In conclusion, we have reported on the temporally and spatially resolved observation of the preliminary stages that lead to the generation of a supercritical unmagnetized plasma shock in a laboratory experiment, in conditions of possible relevance to SNR shocks. A matching simulation allowed us to resolve the long-term evolution of the structure and it provided insight into the physics 
of the transition layer of this supercritical shock. Both experimental data and simulations show that the shock potential has the value necessary to reflect some of the incoming upstream ions which can in turn efficiently heat the foreshock region, thus stabilizing the shock as electrostatic \cite{Forslund,Bardotti,Sorasio,Laming,Treumann}. This effect could reduce the Mach number of the forward shocks of SNRs to values well below the $10^2-10^3$ that one obtains if one estimates their Mach numbers based on the ion acoustic speed of the ISM \cite{Helder2}. 

We acknowledge the contribution of the staff of the Central Laser Facility. We would also like to thank the EPSRC (EP/D043808/1, EP/D06337X/1, EP/I031766/1, EP/I029206/1, EP/F021968/1), Vetenskapsr\aa det (Dnr 2010-4063), the Leverhulme Trust (ECF-2011-383), the Ministry of Education of the Czech Republic (project ECOP Nos. CZ.1.07/2.3.00/20.0279) and ELI-Beamlines CZ.1.05/1.1.00/02.0061), and the Triangle de la Physique RTRA network (ULIMAC) for financial support and HPC2N (Ume\aa) for computer time.

\end{document}